\documentclass[apj]{emulateapj}

\usepackage{color}

\begin{document}

\voffset -1cm

\title{An Unobscured type II quasar candidate: SDSS J012032.19-005501.9}

\author{Y. Li$^{1,2}$, W. Yuan$^1$, H. Y. Zhou$^{3,4,5}$, S. Komossa$^{1,6,7}$, Y. L. Ai$^8$, W. J. Liu$^{3,4,5}$, J. H. Boisvert$^{2}$}

\affil{
$^1$National Astronomical Observatories, Chinese Academy of Sciences, 20A Datun Road, Beijing 100012, China; liye@nao.cas.cn\\
$^2$Department of Physics and Astronomy, University of Nevada, Las Vegas, NV 89154, USA\\
$^3$Key Laboratory for Research in Galaxies and Cosmology, University of Sciences and Technology of China, Chinese Academy of Sciences, Hefei, Anhui 230026, China\\
$^4$Department of Astronomy, University of Science and Technology of China, Hefei, Anhui 230026, China\\
$^5$Polar Research Institute of China, 451 Jinqiao Road, Pudong, Shanghai 200136, China\\
$^6$Excellence Cluster Universe, Technische Universitaet Muenchen, Boltzmannstrasse 2, 85748 Garching, Germany\\
$^7$Max-Planck-Institut fuer Radioastronomie, Auf dem Huegel 69, 53121 Bonn, Germany\\
$^8$Department of Astronomy, Peking University, Beijing 100871, China
}

\begin{abstract}
We report the finding of an unobscured type II Active Galactic Nuclei
(AGN) candidate, SDSS J012032.19-005501.9 
at a relatively high redshift of 0.601, 
which shows a number of unusual properties.
It varies significantly on timescales of years as typical type I AGNs and 
marginally on timescales of weeks. The color-magnitude relation and the 
structure function are also consistent with that of type I AGNs, which 
imply that its variability likely originates from the black hole accretion system. 
However, no broad emission line is detected in the SDSS spectrum, and the 
upper limit of the equivalent width of the H$\rm \beta$ broad emission line
is much less than that of type I AGNs. These properties suggest that
SDSS J012032.19-005501.9 may be an unobscured quasar without broad
emission lines intrinsically, namely an unobscured type II AGN or ``true''
type II AGN. Furthermore, its continuum luminosity 
is at least one order of magnitude fainter than the average value of the
past century expected from the [OIII] emission line.
It indicates that SDSS J012032.19-005501.9 may be switching off.
Additional possible scenarios to explain this intriguing source 
are also discussed.
Future deep observations at multi-wavelengths are needed 
to reveal the nature of this peculiar and intriguing AGN.
\end{abstract}


\section{Introduction}

Active galactic nuclei (AGNs) are observationally classified into type 
I and type II, depending on the presence of broad emission lines (BELs) or lack thereof. 
Such a difference has been successfully explained by different 
viewing angles, known as the unified model of AGNs
\citep{1993ARA&A..31..473A,1995PASP..107..803U}.
Other properties, like the variability can also be explained by this model
\citep{2005ApJ...633..638W,2009AJ....137.5120Y}.
In the unified model, type II AGNs are more edge-on and obscured by the dusty torus, while type I AGNs are more face-on and unobscured. 
The unified model 
is supported by many observations such as the hidden BELs discovered in 
the polarization observation of type II AGNs
\citep{1985ApJ...297..621A,1990ApJ...355..456M,2005AJ....129.1212Z}
and the detection of infrared (IR) BELs
\citep{1994ApJ...422..521G,1997ApJ...477..631V}.

However, the unified model may not be the whole story. Only one half of 
type II AGNs show hidden BELs
\citep{1994ApJ...430..196K,2001ApJ...554L..19T,2007ASPC..373..425M} 
and only 25$\%$ have IR BELs
\citep{1997ApJ...477..631V,2006A&A...457...61R}.
In addition, type II AGNs with significant variability and/or no X-ray 
absorption have been discovered
\citep{2001MNRAS.326..995P,2004A&A...424..519H,2002ApJ...564..196X,2008MNRAS.390.1241B}.
It indicates that there might be AGNs without intrinsic BELs,
since Type II AGNs generally have very high HI column densities
($>$10$^{22} \ \rm cm^{-2}$)
\citep{1999ApJ...522..157R}
and there is no obvious variability for typical type II AGNs
\citep{2009AJ....137.5120Y}.
Such kind of AGNs are called unobscured type II AGNs or ``true'' type II AGNs.
Studies on unobscured type II AGNs may give clues to the formation of the 
broad line regions (BLRs) as well as the formation and evolution of AGNs.

There are a few compelling candidates of unobscured type II AGNs reported,
such as NGC 3147, NGC 4594, NGC 7590, NGC 3660, Q2131-427, Mrk 273x, 
IRAS 01072+4954 and 2XMM J123103.2+110648
\citep{2010ApJ...714..115S, 2012MNRAS.426.3225B, 2012A&A...544A.129V, 2012ApJ...759L..16H}.
Among these candidates, NGC 3147 is the most widely accepted example. This AGN lacks the detection of optical and infrared/polarized BELs, 
and the X-ray/mid-IR observation indicates a low level of obscuration
\citep{2006ApJ...653..127S, 2008MNRAS.385..195B, 2008MNRAS.390.1241B, 2010ApJ...714..115S, 2012A&A...540A.111M}.
The nature of other candidates is still under debate 
\citep{2007ApJ...656..105G, 2009MNRAS.398.1951P,2010ApJ...714..115S,2012MNRAS.426.3225B, 2012A&A...540A.111M,2014AJ....147...12B}.

Observationally, most of the candidates have a low luminosity 
($\lesssim10^{42}$ erg s$^{-1}$)
and a low accretion rate ($\lesssim10^{-3}\ \rm L_{Edd}$)
\citep{2010ApJ...714..115S, 2012MNRAS.426.3225B, 2012A&A...544A.129V, 2012ApJ...759L..16H}.
Both are consistent with the theoretical prediction that the formation of BLRs 
usually requires a high enough accretion rate to maintain a standard 
accretion disk
\citep{2000ApJ...530L..65N, 2003ApJ...590...86L, 2004A&A...428...39C, 
2006ApJ...648L.101E, 2009ApJ...707..233L, 2011MNRAS.417..681L}.
In addition, if the accretion rate is lower and black hole (BH) mass is high,
the emission from the disk would be too weak to provide
sufficient ionizing photons for producing observable broad lines.
Note that IRAS 01072+4954 and 2XMM J123103.2+110648 are two exceptions.
The absence of BELs for these two objects can be explained by the 
high accretion rate and small BH mass.

In this paper, we report another unobscured type II AGN candidate, 
SDSS J012032.19-005501.9. This object was detected as a point source with 
Sloan Digital Sky Survey (SDSS) on September 19th, 1998, and was observed 
repeatedly by the SDSS Supernova Survey, labelled as the SDSS Stripe 82
\citep{2002AJ....123.2945R}.
The follow-up spectrum (mjd-plate-fiberid: 52209-0696-243) shows significant 
narrow emission lines, e.g. H$\beta$ and [OIII] $\lambda5007$, but no BELs.
The redshift was measured to be $z=0.600696\pm0.00019$ at 86\% confidence 
level. 
Several groups originally identified it as a type II AGN
due to emission lines \citep{2003AJ....126.2125Z,2006A&A...455..773V,2008AJ....136.2373R} and/or X-ray/[OIII] luminosity ratio \citep{2010MNRAS.404...48V, 2013ApJ...777...27J}.

This paper is organized as follows. In section 2, data analyses and 
results are presented, with spectrum analyses in section 2.1, 
studies on variability with SDSS Stripe 82 data in section 2.2
and broadband spectrum energy distribution (SED) in section 2.3.
We discuss possible scenarios of its nature and give the conclusion in 
section 3. Throughout this paper, the cosmological parameters $H_0=\, 
70\,\rm km\, s^{-1}\, Mpc^{-1}$, $\Omega_M$=0.3 and $\Omega_{\Lambda}$=0.7 
are adopted.

\section{Data Analyses and Results}
In order to search for unobscured type II AGNs, we used the light-motion 
curves from SDSS Stripe 82 presented in \cite{2008MNRAS.386..887B}.
The Stripe 82 data covers an area of about 300 deg$^2$ in high Galactic 
latitudes, with right ascension (RA) from $\alpha=20^h$ to 4$^h$
and declination (DEC) from $\delta =-1.^{\circ}$27 to $+1.^{\circ}$27.
This region has been observed repeatedly for more than ten years.
In \cite{2008MNRAS.386..887B}, Stripe 82 data from 1998 to 2005 were 
recalibrated for varying photometric zero-points, achieving $\sim 0.02$ 
mag root mean square (RMS) accuracy in the $g$, $r$, $i$ and $z$ bands for 
point sources.

Using this catalog, we first selected objects which are classified 
as ``star'' with photometry in most observations (mean\_obj\_type\_flag$>$5.5)
and required $\chi^2 >$ 3 in both the $g$ and $r$ bands. In total there are 
41,835 objects morphologically classified as ``star''. 
As such, their photometric magnitudes are not influenced by the variable seeing 
and can reflect the variability reliably. Then we cross matched these objects 
with SDSS DR8 catalogue within 3 arcsecs, and selected those with spectral 
classification ``galaxy''. 227 objects are left after this selection.
Finally, we classified them using narrow line ratio \citep{1981PASP...93....5B}
and found that SDSS J012032.19-005501.9 is a type II AGN with significant 
variability. In the following, we present the optical spectrum, variability 
and broadband SED of SDSS J012032.19-005501.9 in detail.

\subsection{Optical Spectrum}

\begin{figure}[!t]
\centering
\includegraphics[width=1.\columnwidth]{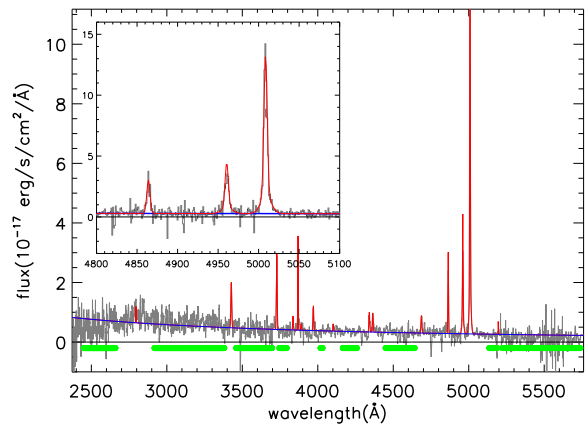}
\caption{Spectrum of SDSS J012032.19-005501.9 (mjd-plate-fiberid: 52209-0696-243).
The grey line is Galactic extinction corrected rest frame spectrum smoothed every 5 pixels. 
The green bands show the regions used for the continuum fit.
The blue line and red lines show the best fitting continuum and emission lines, 
respectively.
Inserted is a zoom-in of the region with H$\beta\lambda$4861, 
[OIII]$\lambda$4959 and [OIII]$\lambda$5007 lines, for a clearer view.
No smoothing is applied.
}
\label{fig:spectrum}
\end{figure}

The spectrum was obtained by SDSS with mjd-plate-fiberid (mpf) 52209-0696-243
and was analyzed with the USTC program \citep{2006ApJS..166..128Z}.
The Galactic extinction corrected rest frame spectrum, smoothed every 5 pixels is shown 
in grey on Fig. \ref{fig:spectrum}
\citep{1999PASP..111...63F}.
The rest frame continuum was fit with a power-law function.
A power law is a good description of the 
continuum due to the low S/N ratio. To avoid the contamination of emission lines and bad pixels, 
we masked the regions around the emission lines as well as the bad pixels.
The spectrum bluer than 2450 $\rm \AA$ was also masked because there are 
many negative values here, which suggests a non-proper 
calibration for this region. The regions used to fit the continuum are marked with 
green bands in Fig. \ref{fig:spectrum}.
The best-fitting result is $F_{\lambda}=(6.8\pm4.2)\times 10^4
\lambda^{-1.46\pm0.07} \times 10^{-17} \rm erg\ s^{-1} cm^{-2} \AA^{-1}$
with a $\chi^2$ by degrees of freedom of $1134/1966$. 
The blue line in Fig.\ref{fig:spectrum} shows this best-fit continuum.
The measured flux of continuum at 5100 $\rm \AA$ is $\rm 
2.38\times 10^{-18} erg\ s^{-1} cm^{-2} \AA^{-1}$. 

After subtracting the power law continuum, we then fitted the 
emission lines in the spectrum. 
We used a single Gaussian function to fit each emission line, except for
[OIII] doublet for which we used double Gaussians. 
The full width half maximum (FWHM) 
of [OIII]$\, \lambda$4959 is tied with that of [OIII]$\, \lambda$5007,
and the flux of [OIII]$\, \lambda$4959 is fixed to be a third of
[OIII]$\, \lambda$5007, which is predicted by photoionization theory.
The best-fit FWHM of [OIII] doublet is $306.4 \pm 15.0$ km/s. 
We plot the best-fitting emission lines in red in Fig. \ref{fig:spectrum}.

To identity whether it is an AGN or starburst galaxy, 
we applied the Baldwin-Phillips-Telervich (BPT) diagram
\citep{1981PASP...93....5B}.
The flux of [OIII] $\, \lambda$5007 is $82.1 \pm 2.6\ \rm erg\ s^{-1} cm^{-2}$,
and the flux of H$\beta$ is $13.6 \pm 1.1\ \rm erg\ s^{-1} cm^{-2}$.
The narrow line ratio [OIII]/H$\beta$ is larger than 6, 
which indicates there is a large chance this object is an AGN. 

\begin{figure}[!thb]
\centering
\includegraphics[width=1.\columnwidth]{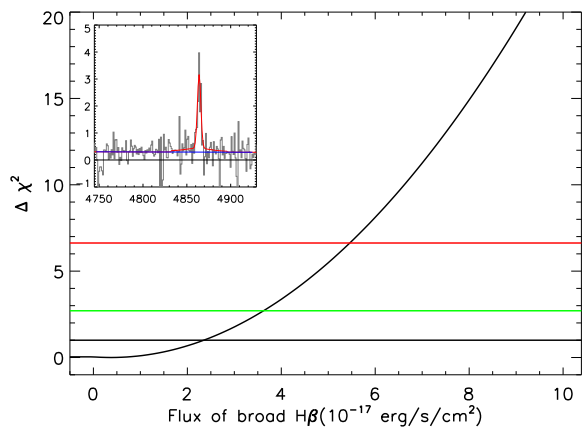}
\caption{Fitting $\chi^2$ versus the total flux of the H$\beta$ BEL. 
The horizontal lines are defined with $\Delta \chi^2=\chi^2-\chi^2_{\rm min}=
1.0,\,2.706$ and $6.63$ from bottom to top, which indicate the 68\%, 90\% 
and 99\% confidence levels respectively. The inserted plot shows the total 
H$\beta$ emission line, with the 90\% upper limit of the BEL added.}
\label{fig:ulimit1}
\end{figure}

To test whether there is a broad H$\beta$ line in the spectrum, we further
added a broad component in the fit. The center of the supposed broad line 
was fixed at 4682.68 $\rm \AA$, and the FWHM was fixed to be 2000 km/s. 
The calculated FWHM is 2156 km/s using $\lambda \rm L_{\lambda}(5100 \rm \AA)$ 
and $M_{\rm BH}$.
$M_{\rm BH}$ was estimated by the FWHM of [OIII] $\lambda$5007 
\citep{2005ApJ...629...61K} due to the lack of broad emissions. 
The total flux of the broad H$\beta$ line was 
assumed to be a free parameter in the fit. No significant broad component 
was found, and the upper limit of the broad line can be derived.
Fig. \ref{fig:ulimit1} shows the $\chi^2$ versus the total flux of the
broad component. The black, red and green horizontal lines indicate the 
68\%, 90\% and 99\% confidence level, defined by $\Delta \chi^2=\chi^2-
\chi^2_{\rm min}=1.0,\,2.706$ and 6.63 respectively 
\citep{1976ApJ...210..642A}.
The $68\%$ and $90\%$ confidence level upper limits of the broad line
were derived to be $2.3\times10^{-17} \rm erg\ s^{-1} cm^{-2}$ and 
$3.6\times10^{-17} \rm erg\ s^{-1}cm^{-2}$. 
The inserted plot in Fig. \ref{fig:ulimit1} shows the total H$\beta$ 
emission line with the 90\% upper limit of the broad component. 

\begin{figure}[!t]
\centering
\includegraphics[width=1.\columnwidth]{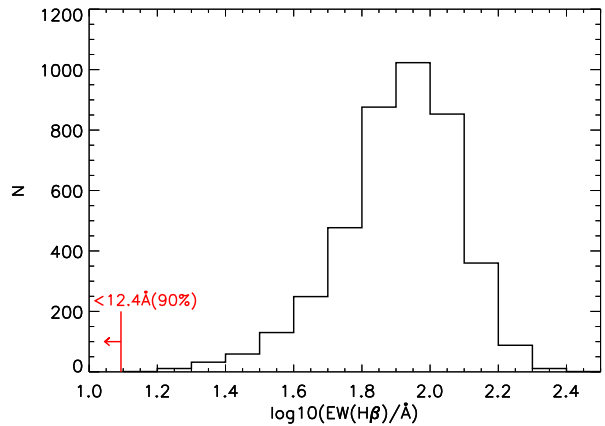}
\caption{Upper limit of the EW of the broad H$\beta$ for SDSS 
J012032.19-005501.9 (red arrow) compared with the EW distribution of 
4171 type I AGNs in \cite{2009ASPC..408...83D}.
}
\label{fig:ewhb}
\end{figure}

The upper limits of the equivalent width (EW) of the H$\beta$ broad line are 
8.0 $\rm \AA$ (68\%) and 12.4 $\rm \AA$ (90\%) respectively. The EW of
broad H$\beta$ line for typical type I AGNs are much larger, as shown by 
the histogram in Fig. \ref{fig:ewhb} 
\citep{2009ASPC..408...83D}. The $90\%$ EW upper limit of SDSS 
J012032.19-005501.9 is smaller than all of the 4171 type I AGN sample in
\cite{2009ASPC..408...83D}. 
Even assuming a broad H$\beta$ line with FWHM 5000 km/s, 
the upper limits of EW are 21.5 $\rm \AA$ (68\%) and 
28.2 $\rm \AA$ (90\%). Only 61 out of 4171 objects have an EW of 
broad H$\rm \beta$ less than 28.2 $\rm \AA$ (1.5\%). 
This result indicates
that there is no BEL in SDSS J012032.19-005501.9 and the spectrum is 
more like a type II AGN.

To evaluate the effect of the absolute flux of the spectrum, we re-calibrated
the raw SDSS spectrum with the photometric data. The photometric data used to 
calibrate the raw spectrum are those obtained on mjd 52207, which is the 
date closest to the time the spectrum was obtained. The photometric magnitudes in the
 $g$, $r$ and $i$ bands on mjd 52207 are $21.64\pm 0.06$ mag, $20.99 \pm 0.04$ mag, 
and $20.18 \pm 0.03$ mag respectively. A wavelength-dependent correction 
factor was adopted since the discrepancies between the photometric data and 
the spectroscopic data vary in different bands. A $\chi^2$ fitting of the 
synthetic magnitudes of the spectrum to the photometric magnitudes resulted 
in a linear calibration relation $F_{\rm cal}/F_{\rm raw}=-1.88+0.000785 
\lambda$ with respect to  wavelength $\lambda$ in observer frame. 

After the calibration, 
the flux of 5100 $\rm \AA$ is $\rm 1.08\times10^{-17} erg\ s^{-1} cm^{-2} 
\AA^{-1}$, corresponding to $L_{\rm bol,5100}=10 \lambda L_{\lambda}(5100)
=8.2\times 10^{44} \rm erg\ s^{-1}$ \citep{2000ApJ...533..631K,
2002A&A...388..771C,2006ApJS..166..470R}.
The luminosity of [OIII] estimated with the raw SDSS spectrum is
$(1.23\pm0.04) \times 10^{42}\ \rm erg\ s^{-1}$
and the calibrated luminosity of [OIII] $\lambda5007$ is $(5.41\pm 0.17)\times 
10^{42}\ \rm erg\ s^{-1}$.
It indicates a bolometric luminosity 
$L_{\rm bol,[OIII]}=3200 L_{\rm [OIII]}= (1.73\pm 0.05) \times 10^{46}\ 
\rm erg\ s^{-1}$ \citep{2006ApJS..166..470R, 2011ApJS..194...45S}, 
which is 20 times larger than $L_{\rm bol,5100}$.

Due to the lack of broad lines and stellar velocity dispersion measurements 
of the host galaxy, the mass of the BH can not be well estimated.
As a very rough estimate, we used the FWHM of [OIII] $\lambda 5007$ as an estimation of the stellar velocity dispersion. 

With an [OIII]$\lambda5007$ FWHM of 306 km/s, the estimated $M_{\rm BH}$ of SDSS J012032.19-005501.9 
is $M_{\rm{BH}}=10^{7.78} M_{\sun} \times (\frac{\rm{FWHM_{[{OIII}}]/2.35}}
{200\ \rm{km/s}})^{3.32} =1.4 \rm \times 10^7\ M_{\sun}$
\citep{2001A&A...377...52W, 2011ApJ...739...28X}.
The scatter of $M_{\rm BH}$ and [OIII] FWHM provides a lower limit of 
the uncertainty of the BH mass, which is $0.5 - 1.0$ dex. 
The Eddington ratio estimated with $L_{\rm bol,[OIII]}$ and $L_{\rm bol,
[5100]}$ is 11.5 and 0.5. The uncertainty of the Eddington ratio is affected 
by the uncertainty of BH mass and should be larger than $0.5 - 1.0$ dex.

\subsection{Optical Variability}

\begin{figure}[!htb]
\centering
\includegraphics[width=\columnwidth]{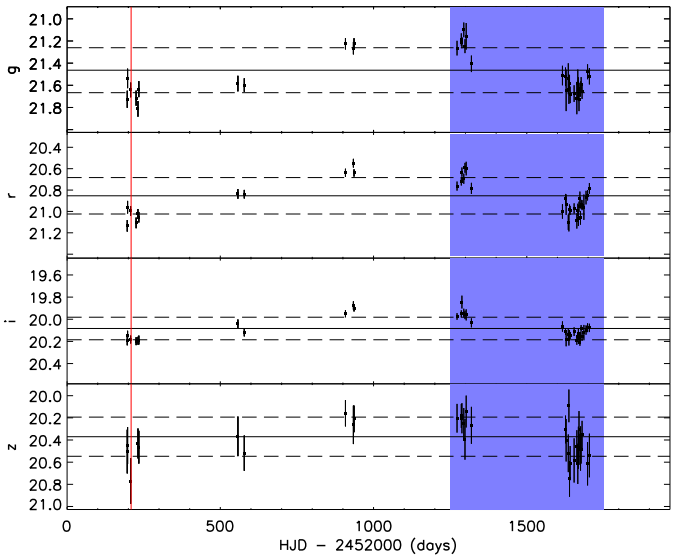}
\caption{Lightcurves of SDSS J012032.19-005501.9 in SDSS photometric bands 
$g$, $r$, $i$ and $z$. In each band the solid line indicates the weighted 
mean value of this band and the dashed lines show the RMS of the points. 
The red vertical line shows the time when the spectrum 52209-696-243 was taken, 
and the blue region marks the period with HJD$-2452000$ from 1250 to 1750.
}
\label{fig:lightcurve}
\end{figure}

For the optical variability of SDSS J012032.19-005501.9, we used the 
Light-Motion Curve Catalogue (LMCC) from \cite{2008MNRAS.386..887B}. 
The $u$ band 
lightcurve was not included because the data quality is very poor. 
There is also one point observed on HJD$-2452000=-924.63779$ not included
in the analysis. The $g$, $r$, $i$ and $z$ band lightcurves are shown
in Fig. \ref{fig:lightcurve}.

The solid black line in each band of Fig. \ref{fig:lightcurve} indicates 
the weighted mean value of this band, defined as $\bar{m}=\frac{\Sigma 
(m_i/\sigma_i^2)}{\Sigma (1/\sigma_i^2)}$, where $m_i$ and $\sigma_i$ are 
the measured magnitude and uncertainty of the $i$th observation. The RMS 
of the data points in each band, $\rm RMS=\sqrt{\frac
{\Sigma (m_i-\bar{m})^2}{N-1}}$, is shown by the dashed lines. We adopted the 
$\chi^2\equiv\Sigma(m_i-\bar{m})^2/\sigma_i^2$ to characterize the variability.
The statistical results of the light curves in the four bands of Fig.\ref{fig:lightcurve} are compiled in Table \ref{table:2}, both for the whole
period with HJD$-2452000=0-1800$ and period $1250-1570$. It is shown that SDSS 
J012032.19-005501.9 varies significantly in the $g$, $r$ and $i$ bands. 
The probabilities of null hypothesis, i.e. no variability, are all less 
than $10^{-16}$ in these three bands. The $z$ band variability is not 
so obvious because of the large photometric uncertainties 
in this band. There seems also to be variability on timescales of weeks, 
though it is only statistically marginal, with null probability $\sim$ 
0.01 in more than one band. The significant variability of SDSS 
J012032.19-005501.9 indicates that it is unlikely to be an obscured type 
II AGN.

\begin{table}[!htb]
\begin{center}
\caption{Statistical results of the optical variability}
\label{table:2}
\begin{tabular}{ccccc|cc}
\hline\hline
band&   mean & RMS &   $\chi^2/{\rm dof}$ & $P_{\rm null}$ & $\chi^2/{\rm dof}$ & $P_{\rm null}$ \\
                  \multicolumn{5}{c}{HJD$-2452000=0-1800$} & \multicolumn{2}{|c}{$1250-1750$} \\\hline
$g$ &      $21.46$    & $0.20$ &         $303.08/35$      &    $  0.00$   &      $169.54/24$           &     $  0.00$     \\
$r$ &      $20.85$    & $0.17$ &         $439.05/37$      &    $  0.00$   &      $205.87/26$           &     $  0.00$     \\
$i$ &      $20.08$    & $0.10$ &         $268.57/37$      &    $  0.00$   &      $133.02/26$           &     $  0.00$     \\
$z$ &      $20.37$    & $0.18$ &         $ 41.74/33$      &    $  0.14$   &      $ 30.39/23$           &     $  0.14$     \\\hline

\end{tabular}
\end{center}
\end{table}


It would be interesting to compare the properties of the variability observed 
in SDSS J012032.19-005501.9 with those detected in normal unobscured AGNs, 
i.e., type I AGNs. Such comparisons may give us hints about
the origin of the variability of SDSS J012032.19-005501.9.

\begin{figure}[!tb]
\centering
\includegraphics[width=\columnwidth]{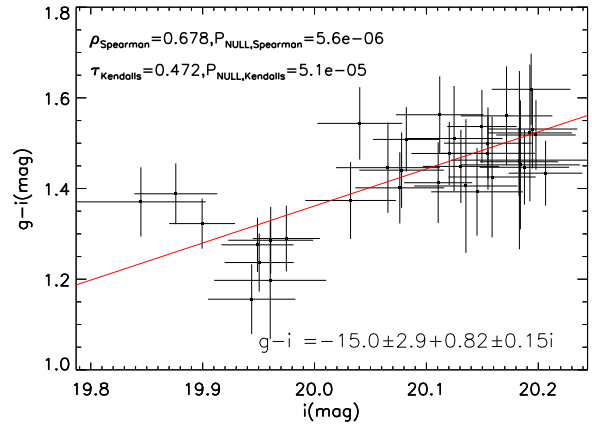}
\caption{Color-magnitude plot of SDSS J012032.19-005501.9.
It shows that SDSS J012032.19-005501.9 becomes bluer when it is brighter.
}
\label{fig:color}
\end{figure}

In general, type I AGNs become bluer when they are brighter
\citep{1990ApJ...354..446W,1999MNRAS.306..637G,2000ApJ...540..652W,
2004ApJ...601..692V}. The theoretical explanations of such behavior 
include a hotter accretion disk, thinner dust absorption,  
contamination of non-variable host galaxies, among others 
\citep{2011ApJ...731...50S}. The color-magnitude plot of SDSS 
J012032.19-005501.9 is shown in Fig. \ref{fig:color}. Here only the 
points with both detections in the $g$ and $i$ bands within 0.1 day are plotted. 
It is 
clear to see that $g-i$ becomes smaller (bluer) when the $i$ magnitude
becomes smaller (brighter). We tested the correlation using Spearman's 
rank correlation. The Spearman's rank correlation coefficient 
$\rho_{_{\rm Spearman}}$ is 0.678 and the null probability 
$P_{_{\rm NULL, Spearman}}$ is $\rm 5.6 \times 10^{-6}$. This result 
indicates that the variability of SDSS J012032.19-005501.9 is similar 
with that of typical type I AGNs, i.e., the brighter the bluer,
although the origin is not clear.

\begin{figure}[!htb]
\centering
\includegraphics[width=\columnwidth]{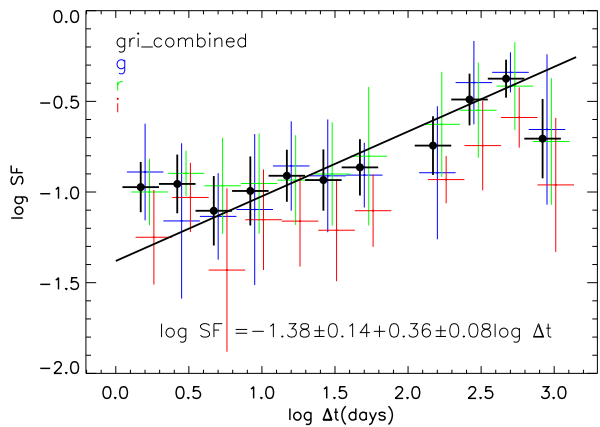}
\caption{The structure function of SDSS J012032.19-005501.9 in SDSS photometric bands 
$g$ (blue), $r$ (green) and $i$ (red), as well as the mean structure function (black).
The results for $g$, $r$ and $i$ bands are shifted rightwards for 
0.05, 0.10, 0.15 dex to have a clear show.
}
\label{fig:5}
\end{figure}

Another commonly used tool to test the variability is the structure function 
\citep{1985ApJ...296...46S,1998ApJ...504..671K,2002MNRAS.329...76H, 
2005AJ....129..615D,2009ApJ...705...46B, 2011A&A...525A..37M, 
1996ApJ...463..466D,2004ApJ...601..692V, 2006AJ....131.1923R,
2008MNRAS.383.1232W}, which shows the variability on timescales ranging 
from days to years. Here we used the modified structure function (SF) defined in 
\cite{1996ApJ...463..466D}
$${\rm SF} (\Delta t)=\sqrt{\frac{\pi}{2}\langle|m(t+\Delta t)-m(t)|\rangle^2
-\langle \sigma_n^2\rangle},$$
where $m(t)$ is the magnitude in time $t$ 
and the terms within the $\langle\rangle$ indicates the mean value.
The first term within the $\langle\rangle$ is the mean of the absolute 
value of the magnitude difference between one pair of points whose time 
lag is $\Delta t$. The $\pi/2$ factor is the Gaussian distribution noise.
$\sigma_n^2$ is the standard deviation in each pair, used to correct for 
the photometric uncertainty
\citep{2008MNRAS.383.1232W}. The time lag $\Delta t$ is computed in 
the rest frame, $\Delta t=\Delta t_{obs}/(1+z)$. The mean of the structure function 
values are calculated for different time bins, with 0.25 dex per bin.
The mean structure function in the $g$ (blue), $r$ (green) and $i$
(red) bands are plotted in Fig. \ref{fig:5}, with $\log \Delta t$ shifting
rightwards for 0.05, 0.10 and 0.15 dex respectively in order to have a
clear view.
The error bar indicates the RMS within that bin, which is much larger than the value obtained from error propagation.
For type I AGNs, the structure function with timescales larger than 10 days can generally 
be well fitted by power law function, with index $\sim0.3-0.4$ 
\citep{2004ApJ...601..692V, 2006AJ....131.1923R, 
2008MNRAS.383.1232W, 2011A&A...525A..37M}. 


After shifting the $r$ ($i$) band structure function by minimizing the $\chi^2$ 
between the $r$ ($i$) and $g$ band structure function values, mean value of the three bands
in each timescale bin was calculated and shown as the black point. 
The uncertainty of this mean structure function was calculated using standard error propagation.
The points with timescales less than 10 days show 
a plateau. This is similar to typical type I AGNs \citep{2013AJ....145...90A}.
The points with timescales larger than 10 days can be fitted by a power-law 
function and the best fitting result of the mean structure function is
$$\log{\rm SF}=(-1.38\pm0.14)+(0.36\pm0.08)\log \Delta t,$$
where the errors are at the 68\% confidence level.
The last point, that shows a drop, 
was not used  because this is a known artifact
of the structure functions  due to the finite length of AGN light curves.
The index is consistent with the result of 
typical type I AGNs
\citep{2008MNRAS.383.1232W, 2011A&A...525A..37M,2013AJ....145...90A}. 
However, it should be noted that the structure functions for individual
AGNs usually  suffer from large statistical uncertainties 
unless their  lightcurves are very well sampled
over a large range of timescales\citep{2010MNRAS.404..931E}.
Cautions should thus be paid when interpreting 
the observed structure function of this object.

In summary, SDSS J012032.19-005501.9 shows significant optical variability 
on timescales of years and marginal variability on timescales of weeks. 
Furthermore, the amplitudes, the color dependence on magnitude,
and the structure function of the variability are all similar to typical type I AGNs.

\subsection{Broadband Spectral Energy Distribution}

\begin{figure}[!b]
\centering
\includegraphics[width=\columnwidth]{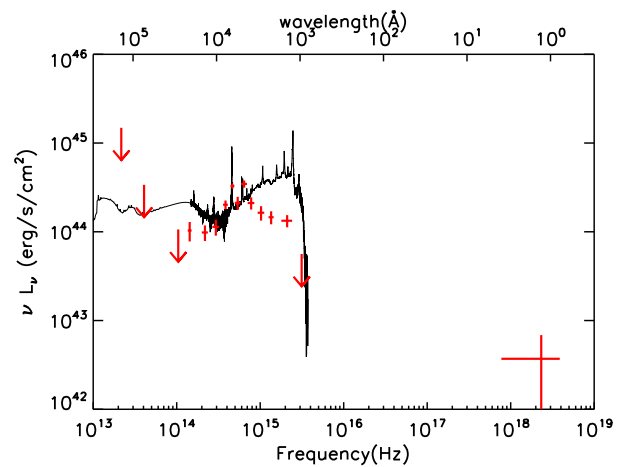}
\caption{Broadband SED of SDSS J012032.19-005501.9 (red crosses and arrows). 
The black line is the composite spectrum of AGNs. A red bump is shown and 
there is no IR bump which is believed to be produced by the dusty torus.
}
\label{fig:sed}
\end{figure}

The multiwavelength data of SDSS J012032.19-005501.9 were compiled from 
the archive database, including the websites of FIRST, WISE, UKIDSS, 
SDSS and Chandra. The SED was calculated correcting for Galactic extinction. If there were multiple observations in
any single band, the weighted mean of all the measurements were adopted.
The errors of these fluxes were adopted to be $max(\rm RMS,\,\sigma_w)$,
where RMS is the scatter between multiple measurements, and 
$\sigma_w\equiv\sqrt{N/\Sigma\frac{1}{\sigma_i^2}}$ is the error of
the weighted mean value which characterizes the measurement uncertainty.
In X-ray band, this source was observed by Chandra  on 02/18/2007, i.e., MJD=54149, and it shows a low X-ray luminosity during that observation
\citep{2010MNRAS.404...48V, 2013ApJ...777...27J}. 
The count rate shown in 
\cite{2010MNRAS.404...48V} is converted to a flux of $3.3^{+2.8}_{-1.6} 
\times 10^{-15} \rm erg\ s^{-1} cm^{-2}$ in $0.5-10$ keV with
WebPIMMs, assuming a power law spectrum 
with Galactic neutral hydrogen $N_{\rm H}$.
The X-ray net counts in 0.5-2 keV and 2-8 keV are $1.9^{+2.6}_{-1.3}$ and
$1.8^{+2.6}_{-1.2}$, respectively.
The rest frame broadband SED is plotted in Fig. \ref{fig:sed} with red 
points. The black line shows the composite spectrum of type I AGN.

The broadband SED peaks at $\sim5 \times 10^{14}$ Hz and shows a sharp 
cutoff in FUV/EUV band. We note that the Big Blue Bump (BBB), which 
peaks around $10^{16}$ Hz and is commonly seen in type I quasars 
\citep{2008ARA&A..46..475H}, is not observed in SDSS J012032.19-005501.9.
The IR bump, which peaks in $\sim 10^{14}$ Hz and is believed to originate 
from the dusty torus, is also absent.

We integrated the broadband SED from optical to X-ray and obtained the 
bolometric luminosity $L_{\rm bol,SED}\approx5.4\times 10^{44}\rm erg\ s^{-1}$.
Such a value is consistent with the bolometric 
luminosity estimated with the optical continuum, $L_{\rm bol,5100}$,
although a little lower.

Assuming the X-ray photon index to be $\Gamma=2.0$, we find
$$\alpha_{ox}=-\frac{\log_{10}[F_{\nu}(2500\AA)]-\log_{10}[F_{\nu}(2\,
{\rm keV})]}{\log_{10}[\nu(2500\AA)]-\log_{10}[\nu(2\,{\rm keV})]}=1.41.$$
This is well consistent with the typical values for type I AGNs
\citep{1998A&A...330..108Y, 2005AJ....130..387S, 2006AJ....131.2826S}, although it is also consistent with the $\alpha_{ox}$ value for absorbed type II AGNs, 
$1.49 \pm 0.16$ \citep{2011MNRAS.416.2792P}.

\section{Discussion and Conclusion}

As mentioned above, the spectrum of SDSS J012032.19-005501.9
lacks detectable BELs, with upper limits on the broad H$\beta$ EW
8.0 $\rm \AA$ (68\%) and 12.4 $\rm \AA$ (90\%). However, its 
variability properties are similar to typical type I AGNs: the 
significant variations, the color-magnitude relation, and the slope of 
the structure function. In addition, the bolometric luminosity of SDSS J012032.19-005501.9
is $L_{bol,5100}=8.2\times10^{44}\ \rm erg\ s^{-1}$, 
which is much lower than the ionizing continuum 
predicted with [OIII] luminosity. 
Now we discuss possible scenarios to explain the 
strange properties of this intriguing source.

\paragraph{(1) A tidal Disruption Event (TDE) in a gas rich environment}

Recently, several candidate TDEs in gas-rich environments have been 
discovered. Their emission-line signature consists of both broad and 
narrow lines, of which broad lines are only detected in the early stages
\citep{2008ApJ...678L..13K, 2011ApJ...740...85W}.
However, if SDSS J012032.19-005501.9 is such an event, it would be 
difficult to understand the optical continuum brightness and the 
variability. In TDEs, the continuum fades away quickly, 
following approximately a $t^{-5/3}$ law. Such a trend has not been seen in 
the optical lightcurve of SDSS J012032.19-005501.9.

\paragraph{(2) An AGN changing its type, from type I to type II, or vice versa}

Several AGNs which change their classifications have been identified. 
Some Seyferts which were previously known as type II have changed to type I,
and vise versa \citep{1971ApJ...164L.109K, 1976ApJ...210L.117T, 2014ApJ...788...48S}. 
However, in  NGC 2617, the change in type was accompanied 
with very high-amplitude optical continuum variability \citep{2014ApJ...788...48S}, 
while the variability of SDSS J012032.19-005501.9 we detect was only mild.

Furthermore, if SDSS J012032.19-005501.9 is such a case, 
it should be in a low state
when it does not show broad emission lines in optical spectrum 
and/or when it shows small X-ray/[OIII] luminosity ratio.
However, a detection of Palomar Transient Factury (PTF) on MJD=54090, 
around 60 days before Chandra observation, shows 20.38 $\pm$ 0.33 mag in V band. 
The data likely indicate that this object was not in a low state 
during the Chandra observation.
It would require that the changes of X-ray and optical continuum 
occur non-simultaneously.
In addition, a neutral hydrogen density $N_{\rm H} > 10^{24} \rm cm^{-2}$, as shown in 
\cite{2013ApJ...777...27J}, is required for a median-high state AGN,
which is also not easy to explain.

However, the uncertainty of PTF detection is very large,
and SDSS J012032.19-005501.9 can still be type 1.9 during Chandra observation.
It can alleviate the difficulty of high $N_{\rm H}$, 
although $N_{\rm H} > 10^{24} \rm cm^{-2}$ is still extreme for type 1.9 AGNs \citep{1999ApJ...522..157R}.
As a result, this scenario cannot be fully excluded, although is not preferred.
If SDSS J012032.19-005501.9 is such a case,  
further spectroscopic monitoring should reveal its change into a type I source at some time.

\paragraph{(3) An AGN with full or partial covering by a dusty absorber}

If a dusty absorber fully covers the continuum source and a major
part of the BLR, then the observed variability timescale of about 1 
year would correspond to the travel time of the dusty absorber.
In addition, the dust should be far enough away from the central continuum 
source in order to survive. With $L_{\rm bol,5100}$ = $8.2 \times 10^{44}\ 
\rm erg\ s^{-1}$ and $L_{\rm bol,[OIII]}=1.73 \times 10^{46}\ \rm erg\ 
s^{-1}$, we expect the dust survival distances to be $0.65 L_{46}^{1/2}\rm\ 
ly=0.18\ ly$ and 0.8 ly \citep{2006LNP...693....1N}, and the required 
velocity of the dusty cloud would be $0.2c$ and $0.8c$, assuming a BH mass 
of $10^7 \rm M_{\sun}$.
If the short timescale variability is true, the required velocity would 
be even larger. Therefore, a full covering absorber is unlikely to explain 
the properties of SDSS J012032.19-005501.9.

Partial covering of the central source and BLR by a dusty absorber can 
provide us with a glimpse into the central engine, of which a certain fraction,
say 10\%, is seen directly. This can explain the observed variability.
Partial covering in the UV has been identified before in several quasars
\citep{1997ApJ...478...80H, 2002ApJ...566..699A}.
Such a scenario can also account for the small [OIII] EW,
provided the partial covering occurs mostly along our line-of-sight.
However, to obscure a significant fraction of the BLR without obscuring 
the central source requires a rather massive and extended absorber with a 
very peculiar geometry. The probability of such a case is relatively small.

\paragraph {(4) An AGN which truly lacks a broad line region.}

In this scenario, SDSS J012032.19-005501.9 is a ``true'' type II quasar, 
i.e. an unobscured type II quasar.

In the unified model of AGNs, the absence of BELs in normal type II AGNs 
is caused by obscuration. However, this is probably not applicable to
SDSS J012032.19-005501.9 because the optical variability of this object
indicates that the optical radiation is more likely emitted from
the central engine directly, presumably the accretion disk, instead of being
scattered as in ``normal'' type II AGNs. In other words, the central
engine of SDSS J012032.19-005501.9 should not be obscured. This
conclusion can be further supported by the analysis of the color-magnitude
relation and the structure function. Like type I AGNs, SDSS J012032.19-005501.9 becomes
bluer when brighter \citep{1990ApJ...354..446W,1999MNRAS.306..637G,
2000ApJ...540..652W,2004ApJ...601..692V}.
Further, its structure function can be fitted with a power law, with an index consistent 
with that of type I AGNs \citep{2004ApJ...601..692V, 2006AJ....131.1923R,
2008MNRAS.383.1232W, 2011A&A...525A..37M}.

Many hints show that the optical variability of AGNs is
driven by fundamental processes, like the accretion rate
\citep{2006ApJ...642...87P, 2008MNRAS.383.1232W}.
The similarity between the variability of this object with typical type I AGNs is likely caused by intrinsic
mechanisms (e.g., an instability or variability in the accretion rate)
rather than a variation of obscuration or extinction.
Furthermore, as shown in \cite{2010MNRAS.404...48V},
the X-ray net counts in 0.5-2 keV and 2-8 keV are $1.9^{+2.6}_{-1.3}$ and 
$1.8^{+2.6}_{-1.2}$ respectively, which are comparable. This indicates 
that the X-ray spectrum may not be as hard as predicted if there is 
significant obscuration of X-rays.

Since we argue that the optical emission of SDSS J012032.19-005501.9 is 
likely intrinsic (not scattered), the very small EW of the BEL, less than 
8.0 $\rm \AA$ (68\%) and 12.4 $\rm \AA$ (90\%), should not be the result of
obscuration --- if any\footnote{A moderate obscuration may exist in
the broadband SED, which can be reproduced by a reddened typical quasar
spectrum with E(B$-$V) $\sim0.1$.}. This is because, in general, the dust 
obscuration is considered to be from the torus, i.e., dusty clouds between 
BLR and NLR. Obscuration from the torus should be the same for both the 
accretion disk and the BLR. Therefore, the extinction of the continuum 
emission and BELs should be similar. But such a scenario cannot explain 
the low EW of the BELs.
We therefore conclude that SDSS J012032.19-005501.9
is a good candidate of a true type II AGN.

\paragraph{(5) An AGN switching off}

The bolometric luminosity of SDSS J012032.19-005501.9 estimated from
the broadband SED and the luminosity of 5100$\rm \AA$ is 
$L_{\rm bol,SED}= 5.1 \times 10^{44}\ \rm erg\ s^{-1}$ and
$L_{\rm bol,5100}=8.2 \times 10^{44}\ \rm erg\ s^{-1}$ respectively. 
However, according to the flux of [OIII]  $\lambda 5007$, the bolometric 
luminosity is $L_{\rm bol,[OIII]}=(1.73\pm0.05) \times 10^{46}\ \rm erg\ s^{-1}$, 
which is more than 20 times higher than $L_{\rm bol,SED}$ and 
$L_{\rm bol,5100}$. The ratio $L_X/L_{\rm [OIII]}$ is 
also smaller by more than one order of magnitude than the typical value 
\citep{2010MNRAS.404...48V, 2013ApJ...777...27J}.


Another possible speculation is SDSS J012032.19-005501.9 may be switching off
\citep{2014arXiv1404.4879D},
or be undergoing large-amplitude variability
\citep{2013AJ....146...78G}.
In the case, the object may have already dropped to a very low state currently, 
in which the broad H$\beta$ line is too weak to be detectable.
The narrow-line emission reflects a previous higher state emission.
Note that the conversion between [OIII] luminosity and the bolometric 
luminosity has a large scatter (about 0.6 dex)
\citep{2006ApJS..166..470R, 2011ApJS..194...45S}, 
and so there is currently no compelling evidence for this scenario.

In summary, the multi-wavelength properties of SDSS J012032.19-005501.9 make
it a good candidate of a true type II AGN, but we cannot yet fully exclude 
the partial obscuration scenario or an AGN changing its type. 
It is also possible that we see an AGN 
switching off. A number of future observations would uncover important 
clues as to the nature of this particular source. For example, deep X-ray 
observations will allow us to search for hard X-rays, present if this galaxy is 
heavily obscured. 
If a faint permanent BLR exists, it may be found with high S/N optical spectroscopy (or polarimetry) and long-term optical spectroscopic 
monitoring could probably detect a spectroscopic type change.

\acknowledgments YL thanks B. F. Liu, D. Xu, Y. Xu and Q. Yuan for helpful discussions.
This work is partially supported by the Natural Sciences Foundation of China
(grant No. 11033007, 11103071 and 11273027)
and the Strategic Priority Research Program of the Chinese Academy of Sciences (grant No. XDB09000000).
H.Z. is supported by Chinese Natural Science Foundation (NSFC-11473025), 
the 973 project (2013CB834905) and Chinese Polar Environment Comprehensive 
Investigation \& Assessment Programmes (CHINARE-2014-02-03).

Funding for the SDSS and SDSS-II has been provided by the Alfred P. Sloan Foundation, the Participating Institutions, the National Science Foundation, the U.S. Department of Energy, the National Aeronautics and Space Administration, the Japanese Monbukagakusho, the Max Planck Society, and the Higher Education Funding Council for England. The SDSS website is http://www.sdss.org/.

The SDSS is managed by the Astrophysical Research Consortium for the Participating Institutions. The Participating Institutions are the American Museum of Natural History, Astrophysical Institute Potsdam, University of Basel, University of Cambridge, Case Western Reserve University, University of Chicago, Drexel University, Fermilab, the Institute for Advanced Study, the Japan Participation Group, Johns Hopkins University, the Joint Institute for Nuclear Astrophysics, the Kavli Institute for Particle Astrophysics and Cosmology, the Korean Scientist Group, the Chinese Academy of Sciences (LAMOST), Los Alamos National Laboratory, the Max-Planck-Institute for Astronomy (MPIA), the Max-Planck-Institute for Astrophysics (MPA), New Mexico State University, Ohio State University, University of Pittsburgh, University of Portsmouth, Princeton University, the United States Naval Observatory, and the University of Washington.


\bibliographystyle{aa}
\bibliography{refs}

\end{document}